\documentclass[useAMS,usenatbib]{mn2e}
\usepackage{booktabs}
\usepackage{amsmath}
\usepackage{amssymb}
\usepackage{graphicx}
\usepackage{cellspace}
\usepackage{dcolumn}
\usepackage{latexsym,graphicx}
\usepackage{color}
\usepackage{hyperref}
\graphicspath{{figures/}}

\newcommand{\MC}{\multicolumn}
\DeclareRobustCommand{\ion}[2]{%
\relax\ifmmode
\ifx\testbx\f
{\mathrm{#1\,\textsc{#2}}}\else
{\mathrm{#1\,\mathsc{#2}}}\fi
\else\textup{#1\,{\mdseries\textsc{#2}}}%
\fi}

\newcommand\kms{\,\text{km}\,\text{s}^{-1}}

\def\apgt{\ {\raise-.5ex\hbox{$\buildrel>\over\sim$}}\ }
\def\aplt{\ {\raise-.5ex\hbox{$\buildrel<\over\sim$}}\ }

\title[Abell\,48 - a rare WN-type CSPN]
      {Abell\,48 - a rare WN-type central star of
        a planetary nebula\footnotemark[0]\thanks{%
Based on observations obtained with the South African Large
Telescope (SALT), commissioning programme \mbox{2010-1-RSA\_OTH-001} and programme \mbox{2011-3-RSA\_OTH-002}.}
}

\author[H. Todt et al.]{%
H. Todt$^{1}$\thanks{E-mail: htodt@astro.physik.uni-potsdam.de}, 
A.\! Y. Kniazev$^{2,3,4}$
V.\! V. Gvaramadze$^{4,5}$,
W.-R. Hamann$^{1}$,
    \newauthor
    D.~Buckley$^{3}$, L.~Crause$^{2}$,
    S.~M.~Crawford$^{2,3}$, A.~A.~S.~Gulbis$^{2,3}$, C.~Hettlage$^{2,3}$,
    \newauthor
    E.~Hooper$^{6}$, T.-O.~Husser$^{7}$, P.~Kotze$^{2,3}$,
    N.~Loaring$^{2,3}$, K. H.~Nordsieck$^{6}$,
    \newauthor
    D.~O'Donoghue$^{3}$, T.~Pickering$^{2,3}$,
    S.~Potter$^{2}$, E.~Romero-Colmenero$^{2,3}$,
    \newauthor
    P.~Vaisanen$^{2,3}$, T.~Williams$^{8}$, M.~Wolf\kern2pt$^{6}$ \\
     $^{1}$University of Potsdam, Institute of Physics and Astronomy,
           14476 Potsdam, Germany\\
     $^{2}$South African Astronomical Observatory, PO Box 9, 7935 Observatory, Cape Town,
             South Africa \\
     $^{3}$Southern African Large Telescope Foundation, PO Box 9, 7935 Observatory, Cape Town,
             South Africa \\
     $^{4}$Sternberg Astronomical Institute, Lomonosov Moscow State University, Moscow, Russia\\
     $^{5}$Isaac Newton Institute of Chile, Moscow Branch, Russia\\
     $^{6}$Department of Astronomy, University of Wisconsin-Madison, 475 N. Charter St., Madison, WI 53706, USA \\
     $^{7}$Institut f\"{u}r Astrophysik Georg-August-Universit\"{a}t, Friedrich-Hund-Platz 1, 37077 Gottingen, Germany \\
     $^{8}$Department of Physics and Astronomy, Rutgers University, 136 Frelinghuysen Road, Piscataway, NJ 08854, USA
}
\begin{document}

\date{Version from { \today }}

\pagerange{\pageref{firstpage}--\pageref{lastpage}} \pubyear{2002}

\maketitle

\label{firstpage}

\begin{abstract}
A considerable fraction of the central stars of planetary nebul\ae\
(CSPNe) are hydrogen-deficient. Almost all of these H-deficient
central stars (CSs) display spectra with strong carbon and
helium lines. Most of them exhibit emission line spectra resembling
those of massive WC stars. Therefore these stars are classed as CSPNe
of spectral type [WC].
Recently, quantitative spectral analysis of
two emission-line CSs, PB\,8 and IC\,4663,
revealed that these stars do not belong to the [WC]
class. Instead PB\,8 has been classified as [WN/WC] type and IC\,4663 as
[WN] type. In this work we report the spectroscopic identification
of another rare [WN] star, the CS of Abell\,48. We performed a
spectral analysis of Abell\,48 with the Potsdam Wolf-Rayet (PoWR)
models for expanding atmospheres. We find that the expanding atmosphere of
Abell\,48 is mainly composed of helium (85 per cent by mass), 
hydrogen (10 per cent), and  nitrogen (5 per cent). The residual
hydrogen and the enhanced nitrogen abundance make this object
different from the other [WN] star IC\,4663. 
We discuss the possible origin of this atmospheric composition.
\end{abstract}

\begin{keywords}
stars: abundances – stars: AGB and post-AGB – stars: mass-loss – stars: Wolf–
Rayet – planetary nebulae: general 
– planetary nebulae: individual: PN~G029.0+00.4
\end{keywords}

\section{Introduction}

During their evolution,
low mass stars in a certain mass range become 
central stars of planetary nebul\ae\ (CSPNe). 
Central stars (CSs) are post-AGB stars with
effective temperatures above 25\,kK, hot enough to ionize
circum-stellar material, which was shed off during the AGB phase. 
The ionized circum-stellar gas becomes visible in the optical as a
planetary nebula (PN) around the CS. 
While most of the low mass stars stay hydrogen-rich at the surface throughout
their life, a fraction of about 20 per cent of CSs
show a hydrogen-deficient surface composition. Almost all of
these H-deficient CSs exhibit spectra with strong carbon and helium
lines, indicating that their surface is composed predominantly by
these elements \citep[e.g.][]{gornytylenda2000,wernerheber1991}.

About half of the H-deficient CSs show emission line spectra resembling
those of massive WC stars, i.e.\ carbon-rich Wolf-Rayet (WR) stars
with strong stellar winds. 
In analogy, the  CSs with bright carbon
emission lines are classified as [WC] stars, where the brackets should
distinguish them from their massive counterparts.

Recently,  quantitative spectral analysis 
of two CSs with broad emission lines 
revealed 
that they do not belong to the [WC] class. The first one was
PB\,8 \citep{todt2010}, which should be classified as [WN/WC] type, 
and the other one was IC\,4663
\citep{miszalski2012}, which shows a [WN] type spectrum. 
In both cases the atmospheres of these stars are mainly
composed of helium.
The atmosphere of PB\,8 also contains hydrogen
and traces of carbon, nitrogen, and oxygen, while IC\,4886
has an almost pure helium atmosphere with only traces of nitrogen.

For some other CSs the membership to the [WN] class was discussed, but
could not be proven \citep{todtpena2010}. 
\cite{depew2011} mentioned the CS of Abell\,48, which
shows a WN-type spectrum, but it was not clear whether this
is really a CSPN and not a massive WN star with a ring nebula. 
In this paper, we present a quantitative analysis of the optical
spectrum of the CS of Abell\,48, which reveals indeed a WN-like 
atmospheric composition.
We will also present arguments for the low-mass nature of Abell\,48.

The paper is organized as follows: Section~\ref{sect:previous} gives an
overview of previous observations of Abell\,48.
In Section~\ref{sect:observation} we
describe our observations and the data reduction.
A short description of our methods is given in Section~\ref{sect:methods}.
The spectral analysis is presented in
Section~\ref{sect:analysis}. Section~\ref{sect:nebula} deals with the
nebula, while
the results and implications are discussed in Section~\ref{sect:discussion}.

\section{Abell\,48}
\label{sect:previous}

\begin{figure*}
\begin{center}
\includegraphics[width=17.5cm,angle=0,clip=]{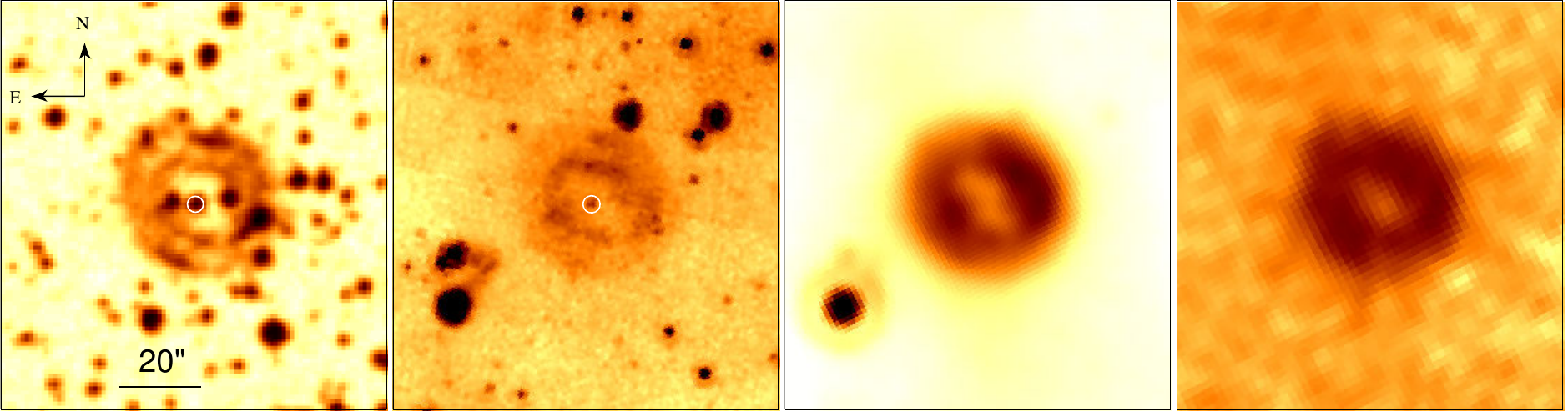}
\end{center}
\caption{
From left to right: DSS-II red band, 
{\it Spitzer} 8 and
24\,$\mu$m, and the VLA 20 cm images of Abell 48 and its
central star (marked by a circle in the DSS-II and 8\,$\mu$m
images). The orientation and the scale of the images are the
same.
    }
\label{fig:neb}
\end{figure*}

Abell\,48 was discovered by \cite{abell1955} on the original plates of
the Palomar Sky Survey. In the red photograph from this survey
\citep[presented in fig.~1 in][]{abell1966},
Abell\,48 appears as a slightly elliptical double-ringed nebula
with a major axis diameter of about $40\arcsec$
(see also the left panel of Fig.~\ref{fig:neb} 
for the Digitized Sky Survey II (DSS-II)
red band \citep{mclean2000} image of the nebula).
\cite{abell1955,abell1966} 
classed the nebula as PN (listed in the discovery papers
under numbers 36 and 48, and named in the SIMBAD database 
PN~A55~36 and PN~A66~48, respectively), while its CS has not been
classified. Since that time, the PN classification of Abell\,48
was generally accepted and the nebula was included in the
Strasbourg-ESO Catalogue of Galactic Planetary Nebulae 
\citep{acker1992} under the name PN~G029.0+00.4.

Abell\,48 is a strong radio source. With the 1.4\,GHz flux of
159$\pm$15\,mJy \citep{condonkaplan1998}, it is one of the brightest
objects among the PNe covered by the NRAO VLA Sky Survey
\citep[NVSS;][]{condoncotton1998}.
The circular shell of Abell\,48 was clearly
resolved in the Multi-Array Galactic Plane Imaging Survey
\citep[MAGPIS;][]{helfandbecker2006},
carried out with the Very Large Array
(VLA)\footnote{The authors of this survey classed Abell\,48 as a
high-probability supernova remnant candidate.}. In the 
MAGPIS~20\,cm image Abell\,48 appears as a thick shell of the same size as
the optical nebula (see Fig.~\ref{fig:neb}).

Abell\,48 was also detected in several major mid- and far-infrared (IR)
surveys, while its ring-like shell has been resolved in the IR
only with the advent of the {\it Spitzer Space Telescope}
 \citep{phillips2008, wachtermauerhan2010}.
Figure~\ref{fig:neb} includes the {\it Spitzer} 8 and 24\,$\mu$m
images of Abell\,48 obtained with the Infrared Array Camera 
\citep[IRAC;][]{fazio2004}
and the Multiband Imaging Photometer for {\it
Spitzer} 
\citep[MIPS;][]{rieke2004}
within the framework of the
Galactic Legacy Infrared Mid-Plane Survey Extraordinaire 
\citep[GLIMPSE;][]{benjamin2003}
and the 24 and 70\,Micron Survey of the Inner
Galactic Disk with MIPS 
\citep[MIPSGAL;][]{carey2009},
respectively. In the 8\,$\mu$m image the nebula shows a bright
inner shell and a diffuse outer halo with an angular extent
equal to that of the outer optical shell, while at 24\,$\mu$m its
morphology is very similar to that of the radio nebula.

To our knowledge, there was not any significant study of Abell\,48
until recently. The interest in Abell\,48 has arisen after 
\cite{wachtermauerhan2010}
carried out optical spectroscopy of its CS and
classified it as WN6. Based on this result, 
\cite{depew2011} and \cite{bojicic2012}
have speculated that Abell\,48 might be another member of the [WN] or
[WN/WC] class \citep{todt2010}, but there has not
been a detailed study yet.

Table\,\ref{tab:det} gives the coordinates and photometry of the CS of
Abell\,48. The $B$ and $R$ magnitudes are from the Guide Star
Catalogue
(GSC2.2)\footnote{\href{http://vizier.u-strasbg.fr/viz-bin/VizieR?-source=I/271}%
{http://vizier.u-strasbg.fr/viz-bin/VizieR?-source=I/271}},
the $I$ magnitude is from the DENIS (Deep Near Infrared Survey of
the Southern Sky) database \citep{denisconsortium2005}, the $J$,
$H$, $K_{\rm s}$ magnitudes are from the 2MASS (Two Micron All Sky
Survey) All-Sky Catalog of Point Sources \citep{cutri2003}, and
the coordinates and the IRAC magnitudes are from the GLIMPSE
Source Catalog (I + II +
3D)\footnote{\href{http://cdsarc.u-strasbg.fr/viz-bin/Cat?II/293}{http://cdsarc.u-strasbg.fr/viz-bin/Cat?II/293}}.
\begin{table}
\begin{center}
\caption{The central star of Abell\,48.}
  \label{tab:det}
  \begin{tabular}{lc}
\toprule
      RA(J2000) & $18^{\rm h} 42^{\rm m} 46\fs91$ \\
      Dec.(J2000) &  $-03\degr 13\arcmin 17\farcs2$ \\
      $l$ & 29\fdg0784 \\
      $b$ & 0\fdg4543 \\
      $B$ (mag) & 19.40$\pm$0.35 \\
      $R$ (mag) & 16.67$\pm$0.34 \\
      $I$ (mag) & 15.50$\pm$0.05 \\
      $J$ (mag) & 13.51$\pm$0.03 \\
      $H$ (mag) & 12.83$\pm$0.03 \\
      $K_{\rm s}$ (mag) & 12.33$\pm$0.03 \\
      $[3.6]$ (mag) & 11.69$\pm$0.06 \\
      $[4.5]$ (mag) & 11.25$\pm$0.10 \\
      $[5.8]$ (mag) & 11.06$\pm$0.09 \\
      $[8.0]$ (mag) & 11.04$\pm$0.16 \\
\bottomrule
    \end{tabular}
\end{center}
\end{table}

\section{Observations and data reduction}
\label{sect:observation}

Abell\,48 was observed within the framework of our ongoing
programme of spectroscopic follow-up observations dedicated to
evolved massive
star candidates that were revealed through the detection of their IR circumstellar nebulae
\citep{gvaramadze2009, gvaramadzefabrika2010a, gvaramadzefabrika2010b,
  gvaramadzehamann2010, gvaramadze2011, gvaramadze2012,
  stringfellow2011, stringfellow2012,burgemeister2013}.
Although Abell\,48 was not included in our
 \citep{gvaramadzefabrika2010a}
list of compact nebulae with central
stars (because it was identified in the SIMBAD database as a PN), we
decided to observe it to check whether its CS is an ordinary WN
star 
\citep[as suggested by][]{wachtermauerhan2010}
or we deal with an
extremely rare object -- a [WN]-type CSPN.

Spectral observations of Abell\,48
were obtained with the Southern African Large Telescope
\citep[SALT;][]{Buck06,Dono06}
on 2011 August 25 during the Performance Verification phase of the
Robert Stobie Spectrograph \citep[RSS;][]{Burgh03,Kobul03},
and one more spectrum was taken on 2012 April 12.
The long-slit spectroscopy mode of the RSS was used,
with a slit width of $1\farcs25$  and a position angle of $0\degr$ 
for all observations.
The seeing was stable in the range of 1\farcs6 to 2\farcs0.
We utilized a binning factor of 2, to give a
final spatial sampling of 0\farcs258 pixel$^{-1}$.
The Volume Phase Holographic (VPH) grating GR900 was used in these observations
to cover a total spectral range of 4300--7300\,\AA\
with a final reciprocal dispersion of $\sim$0.97\,\AA\ pixel$^{-1}$
and a spectral resolution FWHM of 5--6\,\AA.
Three exposures were taken, each with 15 minutes.
Spectra of the Xe comparison arc were obtained to calibrate the wavelength scale.
Spectrophotometric standard stars were observed during twilight.

RSS has three mosaiced 2048$\times$4096 CCDs as a detector.
First, data for each CCD were overscan subtracted, trimmed,
gain and cross-talk corrected, and finally mosaiced.
This primary data reduction was done with the SALT
science pipeline \citep{Crawf10}.
Second, primary corrected and mosaiced long-slit data
were reduced in the way described in \citet{Ketal08}.
Finally, all two-dimensional spectra were averaged.
The spectrum of the surrounding nebula was subtracted using
the IRAF {\tt background} task and a one-dimensional spectrum
of the CS was extracted with the {\tt apall} task.
The two-dimensional spectrum of the surrounding nebula was averaged
along the slit. 

SALT is a telescope with a variable pupil, and its
illuminating beam changes continuously during the
observation. This makes an absolute flux calibration
impossible, even using spectrophotometric standard stars or
photometric standards.
Therefore the spectrophotometric standard stars were used only for a
relative flux calibration, and the absolute flux level was scaled to
the photometric $R$ magnitude.

\section{Methods}
\label{sect:methods}

\subsection{Spectral modelling}

For the analysis of the spectrum 
we used the Potsdam Wolf-Rayet (PoWR) models for expanding
atmospheres \citep[see][]{graefener2002,hamgrae2004}. 
The PoWR code solves the non-LTE radiative transfer
problem in a spherically expanding atmosphere 
simultaneously with the statistical
equilibrium equations while accounting for energy conservation. 
Iron-group line blanketing is treated by 
means of the superlevel approach.
Wind inhomogeneities in first-order approximation
are taken into account, assuming small-scale clumps. 
In this work, we do not calculate hydrodynamically self-consistent
models, 
as in \citet{graefener2005}, but assume that
the velocity field follows a $\beta$-law with $\beta$\,$=$\,1 with
the mass-loss rate as a free parameter. 
Our computations
include complex atomic models for hydrogen, helium,
carbon, nitrogen, oxygen, and the iron-group elements. 
Within our group, this code has been extensively used
for quantitative spectral analyses of massive WR~stars
\citep[e.g.][]{graefener2002, liermann2010} and WR-type CSPN
\citep[e.g.][]{todt2010}.

\subsection{Spectral fitting}
\label{sect:fitting}
The typical emission-line spectra of Wolf-Rayet stars are 
predominantly formed by recombination processes in their 
dense stellar winds. Therefore the continuum-normalized spectrum
shows a useful scale-invariance: for a given stellar temperature
$T_\ast$ and chemical composition, the equivalent widths of the
emission  lines depend only on the ratio
between the volume emission measure of the wind  and the area of the
stellar surface, to a first approximation. 
An equivalent quantity, introduced by 
\citet*{schmutz1989}, is the {\em transformed radius} 
\begin{equation}
 R_\text{t} =
 R_\ast\left[\left. \frac{v_\infty}{2500\,\text{km}\,\text{s}^{-1}}
 \right/ \frac{\dot{M}\sqrt{D}}{10^{-4}\,\text{M}_\odot
 \,\text{yr}^{-1}} \right]^{2/3}~~.  
 \label{eq:abell48-transradius}
\end{equation}
Different combinations of stellar radii $R_\ast$ and mass-loss 
rates $\dot{M}$ can thus lead to nearly the same emission-line strengths. 
In the form given here, the invariance also includes
the micro-clumping parameter $D$, which is defined as the density
contrast between wind clumps and a smooth wind of the same mass-loss
rate. Consequently, mass-loss rates derived empirically
from fitting the emission-line spectrum depend on the adopted value
of $D$. The latter can be constrained by fitting the extended electron
scattering wings of strong emission lines
\citep[e.g.\ ][]{hillier1991,hamann1998}.

\section{Analysis}
\label{sect:analysis}

\subsection{Stellar parameters}
\label{sect:stellarparam}

In order to find the model that best matches to the observation,
we choose a systematic approach:
We begin with a first determination of 
$\log T_\ast$ and $\log R_\text{t}$ with the help of the
already existing model grid for massive WN stars of the late subtypes (WNL)\footnote{%
\href{http://www.astro.physik.uni-potsdam.de/~wrh/PoWR/powrgrid1.html}
{\tt http://www.astro.physik.uni-potsdam.de}} \citep{hamgrae2004}, which
was recently updated with improved atomic data.

For each of the models in this grid, we calculated the reduced $\chi^2_\nu$
\begin{equation} \label{eq:chi2}
 \chi^2_\nu = \frac{1}{N}\sum_{i=1}^N
 \frac{(f_i^\text{obs}-f_i^\text{mod})^2}{\sigma^2} .
\end{equation}
$f^\text{obs}$ denotes the by-eye rectified observed spectrum 
and $f^\text{mod}$ the continuum normalized synthetic spectrum
of the model, each with $N$ data points.  
We consider $R_\text{t}$ and $T_\ast$ as
the fitting parameters. 
We adjusted the uncertainty $\sigma$ so, that the reduced 
$\chi^2_\nu$ of the best fitting model equals unity, corresponding to a mean 
deviation of about 20 per cent. The best fitting model is the one with
$\log T_\ast/\text{K} = 4.85$ and $\log R_\text{t}/R_\odot=0.9$.

The models of this grid had been calculated with a bolometric
luminosity of
$\log L/L_\odot = 5.3$ and a density contrast of $D=4$, as appropriate for
many massive WN stars \citep{hamannliermann2006}.
However, in Section~\ref{sect:discussion} we present arguments that
Abell\,48 is in fact a PN.
Therefore we calculated
a grid of models with parameters that are more appropriate for
CSPNe, i.e.\  for the
stellar luminosity and mass of 
$L=6000\,L_\odot$ and
$M=0.6\,M_\odot$ \citep[see e.g.][]{schoenberner2005,miller-bertolami2007}.

\begin{figure}
\begin{center}
\includegraphics[angle=270,width=0.85\columnwidth]{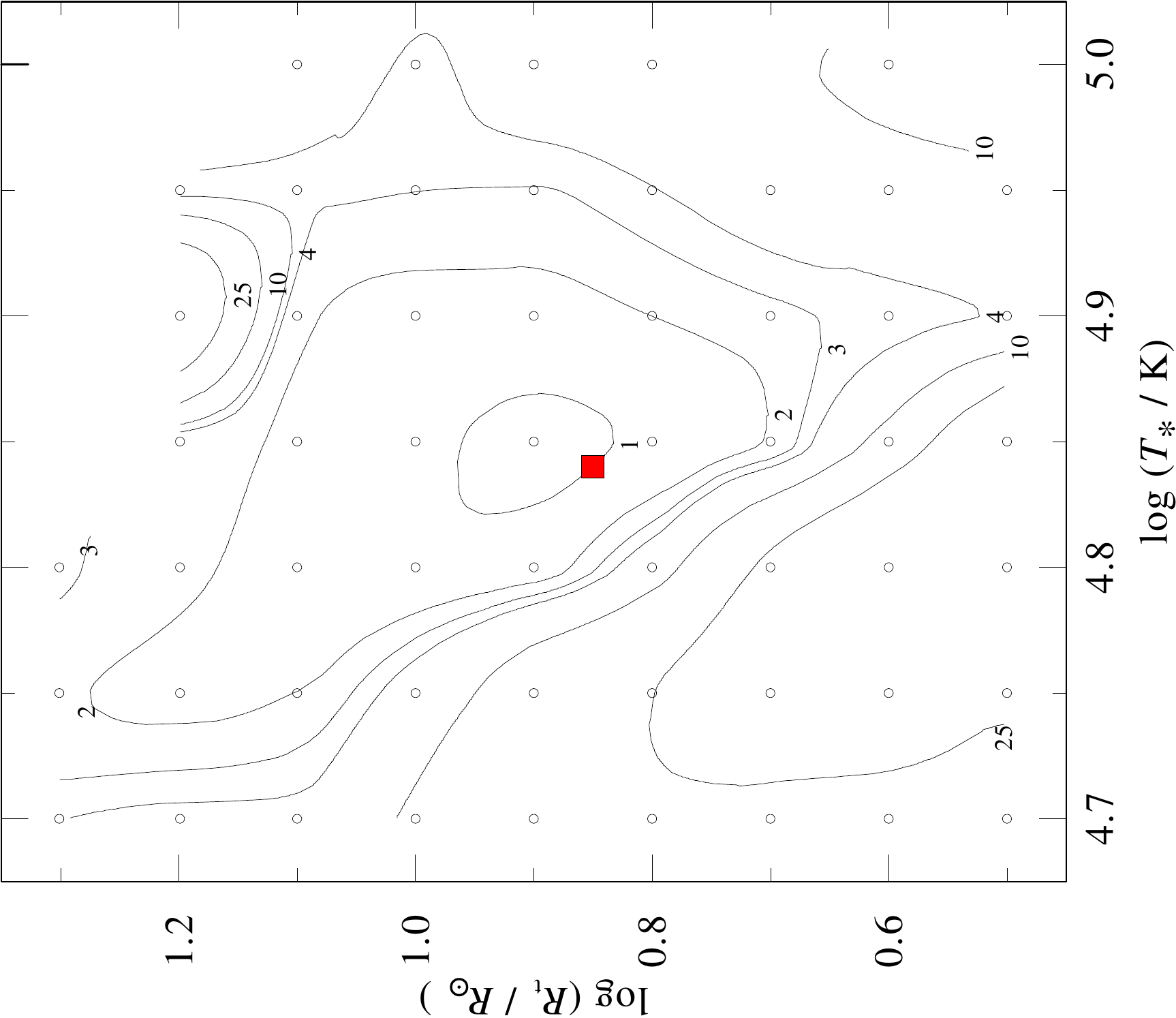}
\caption{Contours of the reduced $\chi^2_\nu$ 
for our grid of [WN] models. 
The open circles indicate the calculated models. 
 Between these data
 points the contour lines are interpolated.
Note that a $\Delta \chi^2_\nu =1$
corresponds to a $1\sigma$ confidence interval. Thus, the contour with $\chi^2_\nu=2$ 
indicates the $1\sigma$ uncertainty of the parameters $R_\text{t}$ and 
$T_\ast$. Our final model is indicated by the red square.} 
\label{fig:chi-kontur-wnlcs}
\end{center}
\end{figure} 
%
%
\begin{figure}
 \includegraphics[width=0.99\columnwidth]{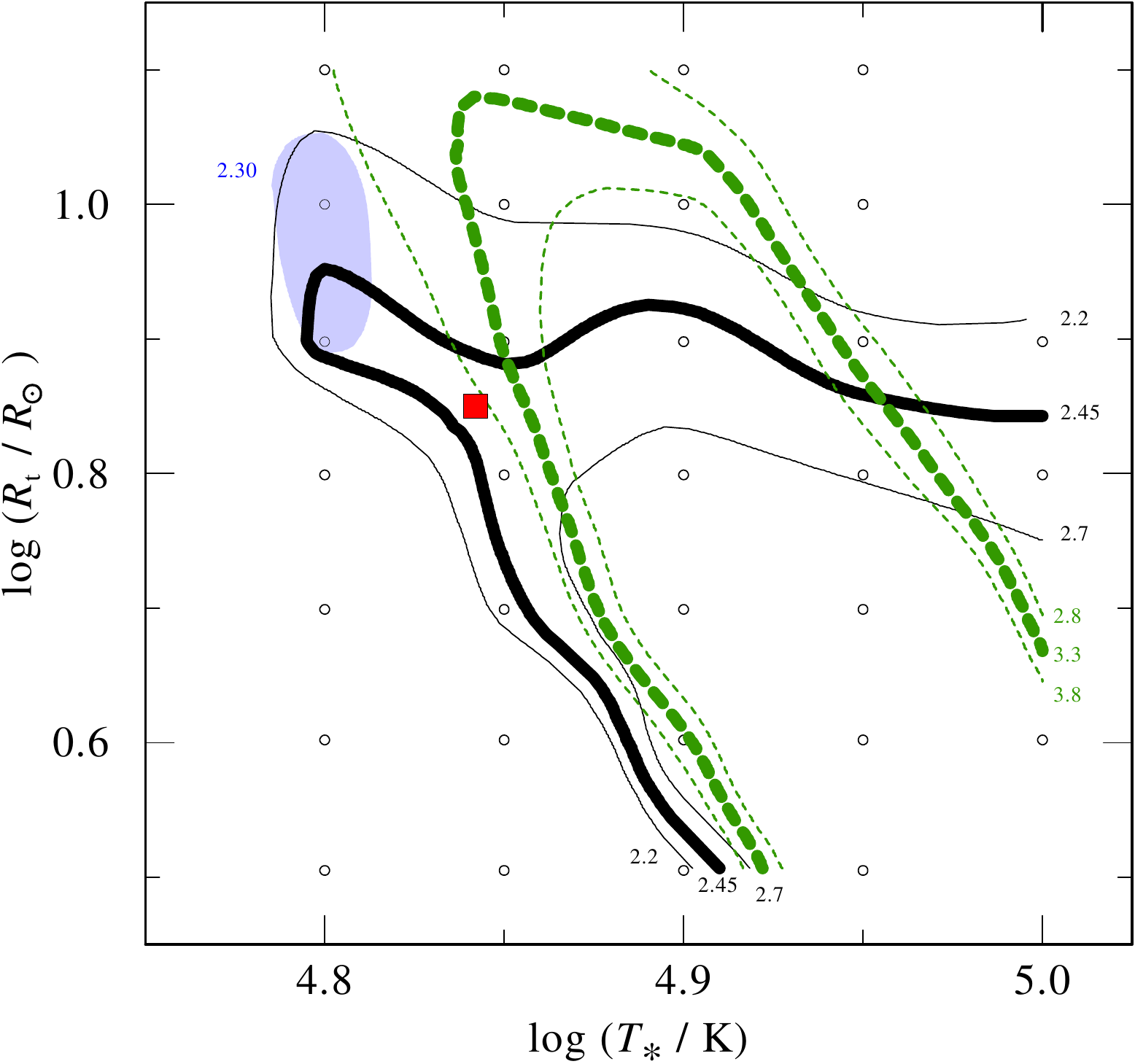}
 \caption{
Contours of the ratio between the line peak heights
of 
N\,{\sc iv} 7100 / N\,{\sc iii} 4643 (green dashed lines), 
N\,{\sc iv} 7100 / N\,{\sc v} 4933 (blue filled ellipse), and
He\,{\sc ii} 5412 / He\,{\sc i} 5876 (black solid lines).
The thick contours represent the measured values, the thinner lines
the 1$\sigma$ uncertainty.
The other symbols are the same as in
Fig.~\ref{fig:chi-kontur-wnlcs}.}
\label{fig:iso_peak_all}
\end{figure}

Moreover, with our first estimate of $R_\text{t}$ and $T_\ast$ we
calculated a series of models with different chemical abundances to determine
the most appropriate atmospheric composition. 
The best fit is achieved with the 
abundances presented in Table~\ref{tab:parameters}, as described in
Section~\ref{subsect:elements}. These abundances were 
used for our [WN] grid. 

Since there is no  UV observation available 
that would contain spectral lines with 
P Cygni profiles, we had to deduce the wind terminal velocity,
$v_\infty$ from the
width of the optical emission lines. The best fit is
achieved with $v_\infty=1000\pm200\,$km\,s$^{-1}$. 
Line broadening by microturbulence is also
included in our models. From the shape of the line profiles we estimated
the microturbulence to be about $100\,$km\,s$^{-1}$.

We repeated our $\chi^2_\nu$-fit analysis for
this grid of [WN] models to get
the uncertainties in $R_\text{t}$ and $T_\ast$ 
(see Fig.~\ref{fig:chi-kontur-wnlcs}). The 
$1\sigma$-uncertainties can be obtained
from the contour with $\chi^2_\nu=2$, i.e. where the value of  
$\chi^2_\nu$ has increased by $1$ compared to the best-fitting model, 
for which $\chi^2_\nu=1$. The best-fitting model from this [WN] grid, 
has $T_\ast = 71^{+12}_{-\hphantom{0}9}\,$kK 
and $\log R_\text{t}/R_\odot = 0.9 \pm0.2$,
where the given error margin corresponds to a doubling of the
deviation between model and observation, as measured by the reduced
$\chi^2_\nu$ (see Equation~\ref{eq:chi2}).

For a further refinement, we calculated 
the line-strength ratios between significant lines of
He{\sc\,ii}\,/\,He{\sc\,i}, N{\sc\,iv}\,/\,N{\sc\,iii}, 
and N{\sc\,iv}\,/N{\sc\,v} for this grid and compared them to
the observation (cf.\ Table~\ref{tab:abell48-peakratios}).
We used line ratios instead of absolute line strengths to
diminish the influence of chemical abundances. In practice, it is often
difficult to measure the equivalent widths of lines correctly, e.g.
due to line blends. Therefore we used the peak height as a measure of
the line strength.

From our optical observation we estimate the uncertainty
in the normalized continuum to be of the order of 
10 per cent in the region of the N{\sc\,iii}\,$\lambda$4643 line 
and 5 per cent at the other diagnostic lines.
We consider this as the uncertainty in the peak measurement, which we then
use to infer a 10 per cent uncertainty in the measured 
 He{\sc\,ii}\,/\,He{\sc\,i}
 and 
 N{\sc\,iv}\,/N{\sc\,v} peak height ratio  and a 15 per cent uncertainty
in the measured  N{\sc\,iv}\,/\,N{\sc\,iii} peak height ratio. 

In Figure~\ref{fig:iso_peak_all} we show contours of the
N{\sc\,iv}\,/\,N{\sc\,iii} and He{\sc\,ii}\,/\,He{\sc\,i} line
peak ratios.
The bold lines correspond to the observed values.
These contours have two intersections.
As a further criterion we now consider the N{\sc\,iv}\,/N{\sc\,v} ratio.
Unfortunately, all models 
of the grid give a smaller value than observed, but the two models
encircled by the shaded area in the upper left corner of the diagram
come closest. 
After manual inspection of the spectral fits we finally decided for
 the final model with 
$\log R_\text{t}/R_\odot = 0.85^{+0.15}_{-0.13}$ and
$T_\ast = 70^{+4}_{-5}$\,kK,
indicated by the red square in Figure~\ref{fig:iso_peak_all}.
The given uncertainties were estimated from the 1$\sigma$ contours in
this figure.

Remarkably, the final parameters derived under assumption of a CS
luminosity hardly differ from those obtained above from the massive
star grid. This demonstrates that the scale-invariance for models 
with same $R_\text{t}$ holds over a wide parameter range.
 
Together with the normalized spectrum we obtained the synthetic
spectral energy distribution (SED) in absolute units from the PoWR
models. The synthetic SED was fitted to the flux-calibrated spectrum 
and photometric measurements by adjusting the distance and the
reddening parameter $E_{B-V}$ (cf.\ Fig.~\ref{fig:abell48_sed}).
The photometric magnitudes for the Johnson and 2MASS measurements
were converted to fluxes with help of the
calibrations by \cite{holbergbergeron2006}, the IRAC measurements with
help of the calibrations by \cite{cohen2003}.

The best SED-fit was obtained with a color excess of $E_{B-V} =
2.10$\,mag and the standard total-to-selective absorption ratio $R_V=3.1$
\citep{fitzpatrick1999}. We estimated a distance of $d=1.9$\,kpc
towards Abell\,48 with the adopted stellar luminosity of
$6000\,L_\odot$.

%

\begin{table}
\begin{center}
\caption{
 Ratios between the peak heights: measured ratios, inferred quantities, 
 and ratios from the final model. 
\label{tab:abell48-peakratios}
}
\begin{tabular}{lr @{$\pm$} lc}
\toprule
Line ratio & \multicolumn{2}{c}{Observed} & \multicolumn{1}{c}{Final model} \\
\midrule
He\,{\sc ii}\,$\lambda$5412 / He\,{\sc i}\,$\lambda$5876 & $2.45$ & $0.25$ 
                           &  2.2  \\[0.1cm]
N\,{\sc iv}\,$\lambda$7120 / N\,{\sc iii}\,$\lambda$4634 & $3.3$ & $0.5$ 
                           & 2.3  \\[0.1cm]
N\,{\sc iv}\,$\lambda$7120  / N\,{\sc v}\,$\lambda\lambda$4933/43 &
$2.8$ & $0.3$ & 2.1 \\     
\bottomrule
\end{tabular}
\end{center}
\end{table}

%
\begin{figure*}
 \includegraphics[width=\textwidth]{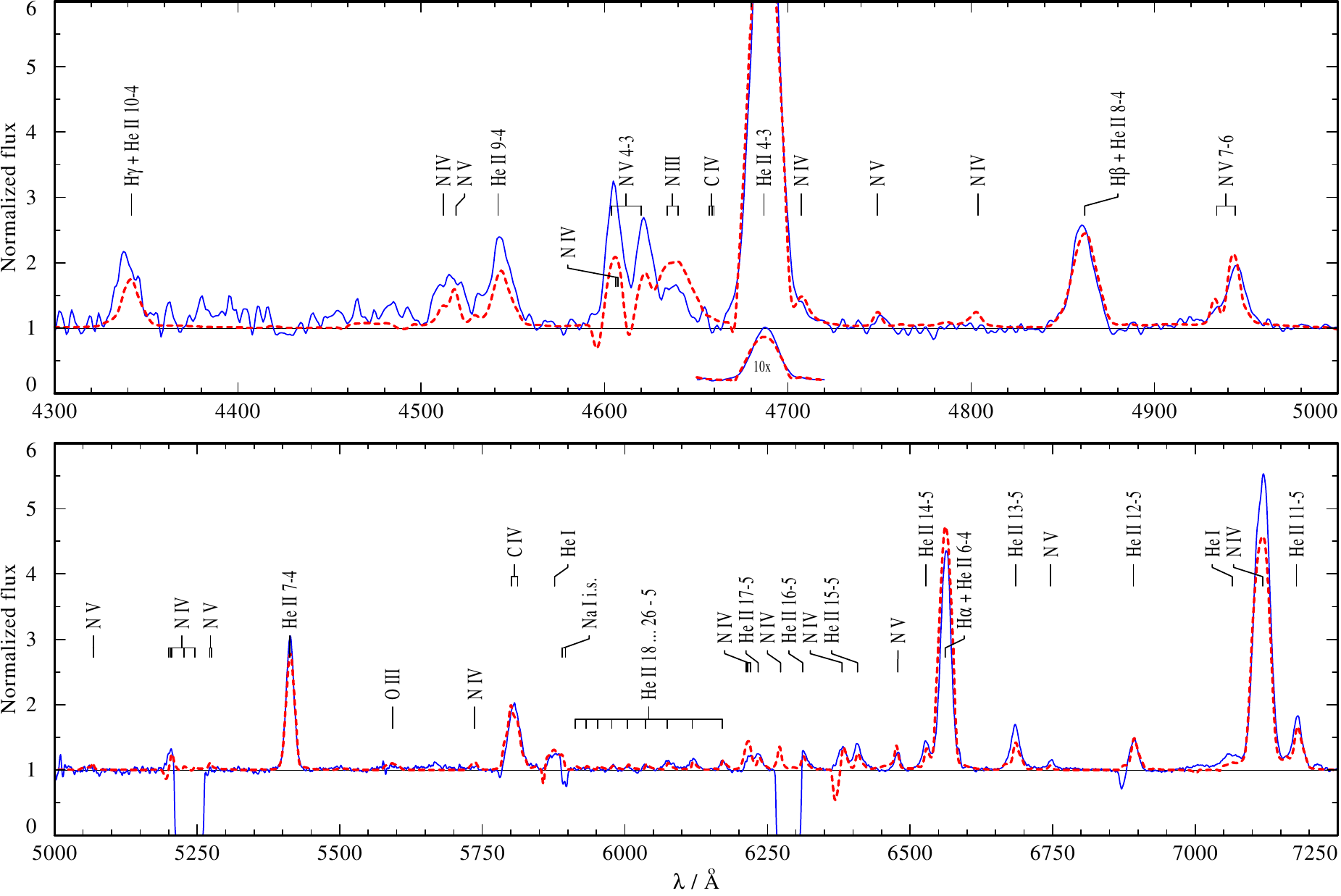}
 \caption{Optical spectrum of the CS of Abell 48: Observation (blue solid) vs.\ 
          synthetic spectrum (red dashed). The observed spectrum has
          some gaps and gets much noisier towards shorter wavelengths.}
\label{fig:spec_norm}
\end{figure*}
%
\begin{figure*}
 \includegraphics[width=\textwidth]{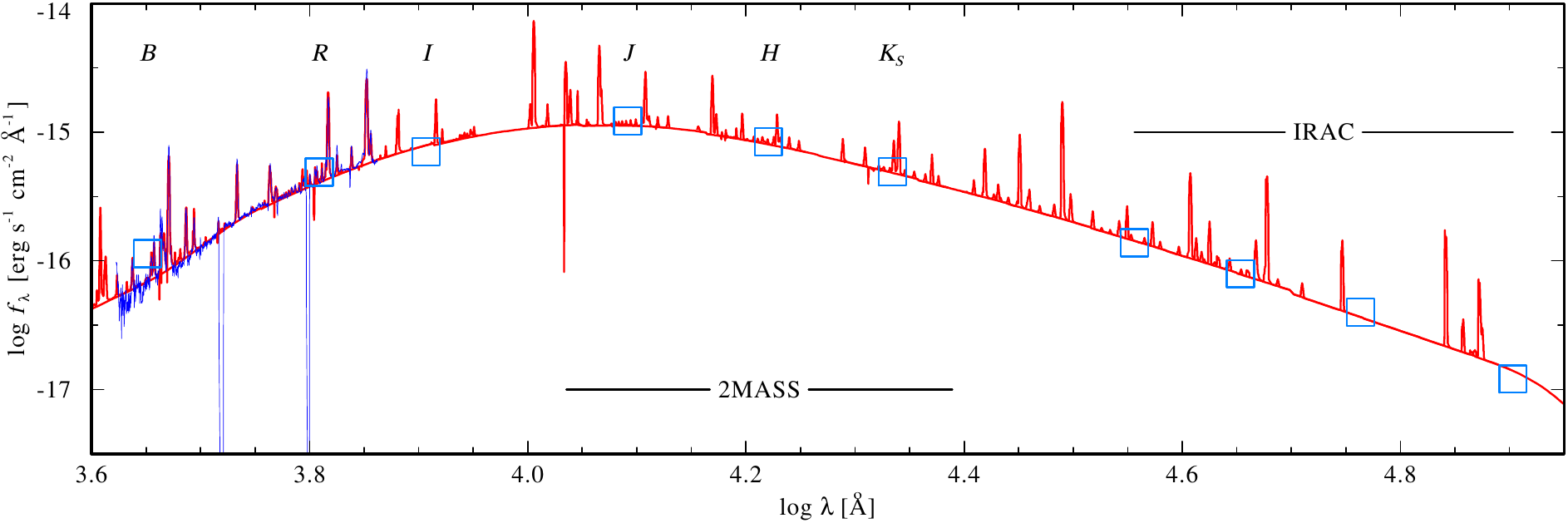}
 \caption{Observed flux distribution of Abell\,48 (blue/noisy) in
 absolute units, including the calibrated spectrum and the photometric
 measurements (open blue squares, see Table~\ref{tab:det}), compared to the emergent
 flux of the model (red/smooth line). 
 The model flux has been reddened
 and scaled to the distance according to the parameters given in 
 Table~\ref{tab:parameters}.
\label{fig:abell48_sed}
}
\end{figure*}

\subsection{Element abundances}
\label{subsect:elements}

In an iterative process with the determination of
$T_*$ and $R_\text{t}$, 
we varied the abundances of He, H, C, N, and O. 
As there is no UV-spectrum of the CS of Abell\,48 available that would help to
determine the abundances of the iron-group elements, we
kept latter fixed to solar values. 

{\em Helium}. 
From the strengths of the helium emission lines, especially 
He\,{\sc ii}~4-3, it is already obvious that the wind 
is mostly composed of helium.

\begin{figure*}
 \includegraphics[width=0.8\textwidth]{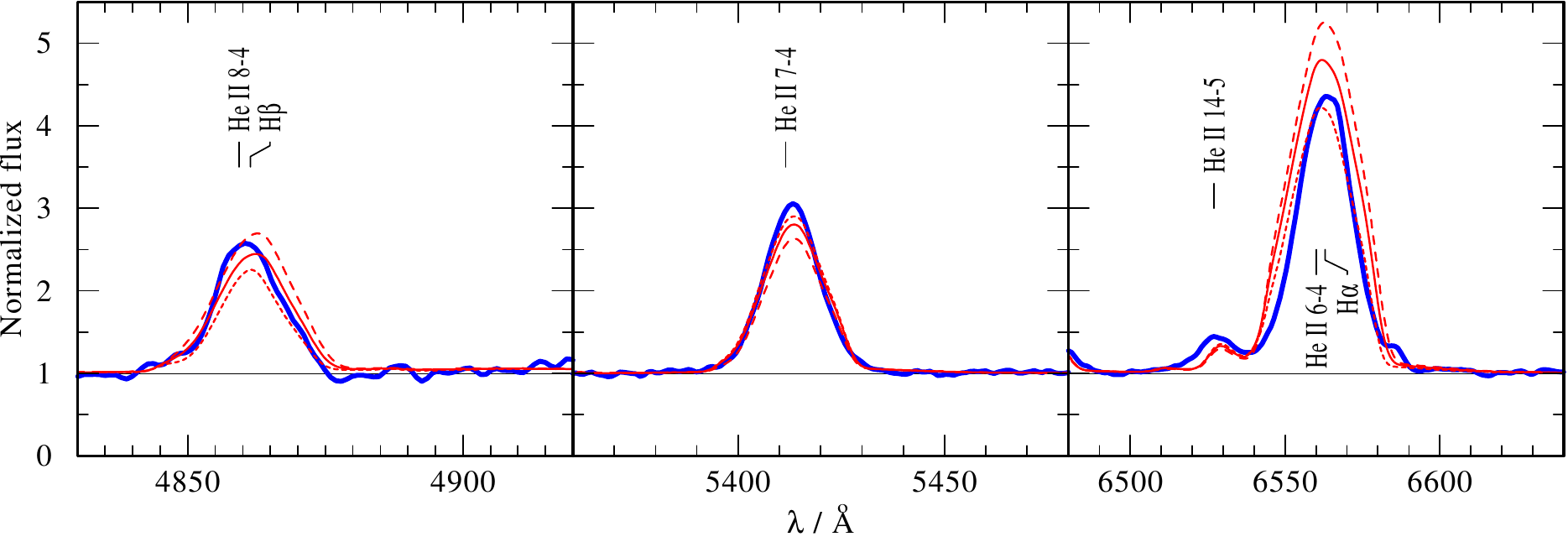}
 \caption{
Test for the hydrogen abundance in the CS of Abell\,48. The observed profiles
(blue/thick solid line) are compared with models (red/thin 
lines) for different hydrogen abundance, but otherwise the same
parameters as of our best-fit model (see Table~\ref{tab:parameters}). 
The tested hydrogen 
mass fractions are zero (dotted lines), 10 per cent (thin solid lines)
and 20 per cent (dashed lines), respectively. 
}
\label{fig:hydrogen}
\end{figure*}

{\em Hydrogen}. 
None of our models is able to reproduce the H$\alpha$/He\,{\sc ii} 
and H$\beta$/He\,{\sc ii} blends
together with the unblended He\,{\sc ii}~7-4 line simultaneously as
observed (cf.~Fig.~\ref{fig:hydrogen}).
H$\beta$ fits best with a hydrogen mass fraction of 20 per cent, while 
the H$\alpha$/He\,{\sc ii} blend matches the observation with about 3
per cent of hydrogen.
As a compromise we adopt a value of 10 per cent of hydrogen.
A better resolved 
spectrum with a higher signal-to-noise ratio (S/N) would help to pin
down the exact hydrogen abundance.

{\em Nitrogen}. 
The nitrogen lines in the spectrum of the CS of Abell\,48 are
much stronger than in WN spectra of stars with similar parameters.
We had to increase the nitrogen abundance to 5 per cent by mass to match
the observed line strength of the N\,{\sc v}~$\lambda\lambda$4933/4944
multiplet and most of the other nitrogen lines. The N\,{\sc iv}~$\lambda$7125
line and the  
N\,{\sc iv}/N\,{\sc v} blend at 4600\,\AA\ are unfortunately not well
matched by our model. Models with only $X_\text{N}=3$ per cent provide
a much better fit to the N\,{\sc iii}\,$\lambda$4634 line, while the strength
of the N\,{\sc iv}~$\lambda$7125 line is best reproduced by models with 
$X_\text{N}=7$ per cent. Generally, it is hard to find a sufficient fit
for spectral lines of three different ions from the same element. This might be
an indication for an inhomogeneous wind. 

{\em Carbon}. 
In the absence of other carbon lines within the range of our
observed spectrum we have to rely on the C\,{\sc iv}~$\lambda$5800 line only.   
This line can be best fitted with a model that has a carbon abundance of
0.3 per cent by mass. The given uncertainty of $\pm 0.1$ per cent accounts only for
the uncertainty of the continuum normalization 

{\em Oxygen}. There are no strong oxygen lines in
the spectral range of our observation. A model with solar 
oxygen abundance of $X_\text{O}=0.6$ per cent 
by mass predicts a weak emission line 
from O{\sc\,iii}~$\lambda$5592 with a line strength that is almost below the 
detection limit. A higher oxygen abundance would not be compatible with 
the observation.

The final line fit is shown in Fig.~\ref{fig:spec_norm} and the
model parameters are compiled in Table~\ref{tab:parameters}. 

\begin{table}
\centering
\caption{Abell 48: Stellar and wind parameters.}
\label{tab:parameters}
\begin{tabular}{SlSrSl}
\toprule
$T_*$                   & $70\,^{+4}_{-5}$               & $\text{kK}$\\
$v_\infty$              & $1000\pm 200$           & $\text{km}\,\text{s}^{-1}$\\
$\log R_\text{t}$       & $0.85\,^{+0.15}_{-0.13}$  & $R_\odot $\\
$\log \dot{M}$          & $-6.4\pm 0.2$ & $M_\odot\,\text{yr}^{-1}$\\
$R_\ast$                & $0.54\pm 0.07$   & $R_\odot$\\
$D$                     & $4$       & (density contrast)\\
$E_{B-V}$ & $2.10\pm0.05$         & $\text{mag}$\\
$v_\text{rad}$     & $50\pm4$                  & $\text{km}\, \text{s}^{-1}$\\
H                  & $10 \pm 10$    & per cent by mass   \\
He                 & $85 \pm 10$    & per cent by mass       \\
C                  & $0.3 \pm 0.1$  & per cent by mass       \\
N                  & $5.0 \pm 2.0$  & per cent by mass      \\
O                  & $\leq 0.6$    & per cent by mass \\
Fe                 & $1.4\times 10^{-3}$ & per cent by mass     \\
\bottomrule
\end{tabular}\\[.3cm]
\end{table}

\section{The nebula of Abell\,48}
\label{sect:nebula}

\begin{figure}
 \includegraphics[angle=0,width=0.9\columnwidth]{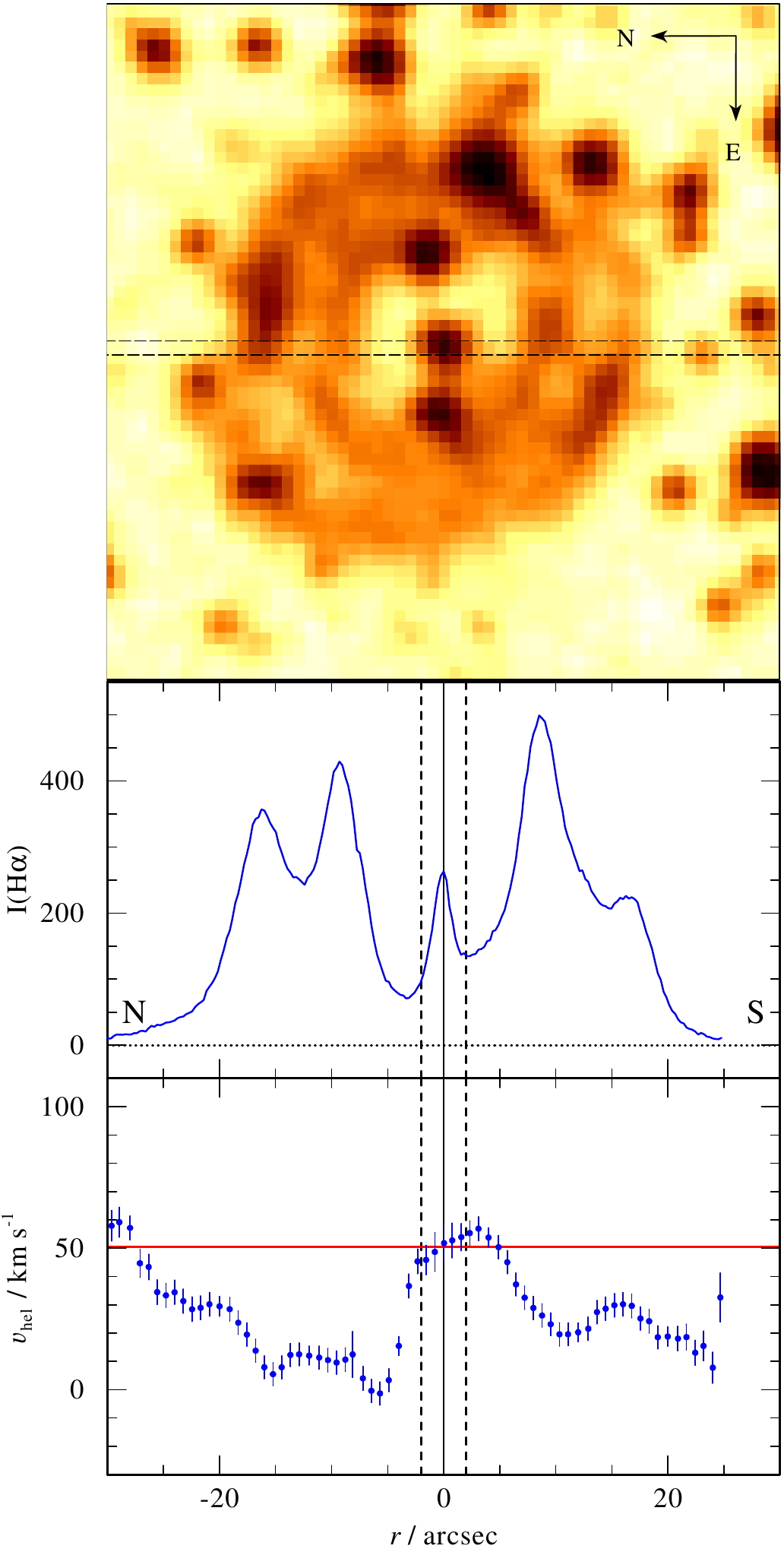}
 \caption{
H$\alpha$ intensity and radial velocity distribution
along the slit.
{\it Upper panel:} DSS-II red band image of Abell\,48 with
the position of the $1\farcs25$ slit shown by dashed lines.
{\it Middle panel:}
The relative flux of the H$\alpha$-line along the slit
after continuum subtraction. N-S direction of the slit is shown.
The vertical line at $r=0$ corresponds to the position of the CS.
The region $r\pm2\arcsec$ is shown with vertical, dashed lines
to mark the area, where the average velocity
$v_\text{hel} = 50.4 \pm 4.2$\,km\,s$^{-1}$ was calculated.
{\it Bottom panel:} The velocity profile of the
H$\alpha$ line, corrected by using the night-sky line [\ion{O}{i}]
$\lambda$\,6363\,\AA. The horizontal line indicates the calculated
velocity $50.4$\,km\,s$^{-1}$, that has to be close to the
heliocentric velocity of the CS.
\label{fig:PN_Vel}}
\end{figure}

\subsection{Position--velocity diagram along the slit}
\label{txt:PV_body}

In  Fig.~\ref{fig:PN_Vel} we present the H$\alpha$ intensity and
radial velocity distributions along the slit, and the
DSS-II red band image of Abell\,48 for comparison.
The extension of the nebula in the H$\alpha$ line is very
similar to that in the DSS-II image. H$\alpha$ emission
appears everywhere within the nebula and its peaks are
clearly correlated with the CS and the shell's position.
We measured the diameter of the nebula of 43$\arcsec$ at a ten per
cent level of the peak intensity in the H$\alpha$ line.
At the distance of Abell\,48 of $1.9$\,kpc (see
  Sections~\ref{sect:stellarparam} and \ref{sect:location})
the radius of the nebula is about 0.2\,pc.

We calculated the line-of-sight velocity distribution ($v_\text{hel}$) 
along the slit
using the method and programs described in \citet{Zasov00}.
To exclude systematic shifts originating from 
the known RSS flexure,
the closest bright night sky line [\ion{O}{i}] $\lambda$6363\,\AA\ was used.
Finally, only those velocity estimates where used which
satisfy the criteria S/N~$>$~3 and $\sigma_v$~$<$ 10\,km\,s$^{-1}$.

Fig.~\ref{fig:PN_Vel} shows that
$v_{\rm hel}$ along the slit spans a rage of $\approx 0$
to $50\kms$. This range is very similar to that found
for PNe \citep[see e.g.,][]{TAGR09}.
We measured $v_\text{hel} = 50.4 \pm 4.2$\,km\,s$^{-1}$ in the area which is
closest to the position of the CS and suggest that this velocity
corresponds to the velocity of the mass center.
With our spectral resolution ($\text{FWHM}=5.5\pm0.5$\,\AA)
for the range of the H$\alpha$ line, we are not able to measure 
the expansion velocity
directly.

\subsection{Physical conditions and chemical abundances}
\label{sect:Phys}

\begin{table}
\begin{center}
\caption{Line intensities of the nebula Abell\,48.}
\label{tab:Intens}
\begin{tabular}{lr@{$\pm$}lr@{$\pm$}l} 
\toprule
$\lambda_{0}$(\AA) Ion    & \multicolumn{2}{c}{F($\lambda$)/F(H$\beta$)}
& \multicolumn{2}{c}{I($\lambda$)/I(H$\beta$)}\\
\midrule
4340\ H$\gamma$\          & 0.170 & 0.009 & 0.451 &0.024 \\
4363\ [O\ {\sc iii}]\     & 0.014 & 0.005 & 0.034 & 0.013 \\
4861\ H$\beta$\           & 1.000 & 0.031 & 1.000 & 0.033 \\
4959\ [O\ {\sc iii}]\     & 1.187 & 0.030 & 1.005 & 0.026 \\
5007\ [O\ {\sc iii}]\     & 4.050 & 0.094 & 3.165 & 0.077 \\
5755\ [N\ {\sc ii}]\      & 0.018 & 0.002 & 0.004 & 0.001 \\
5876\ He\ {\sc i}\        & 0.974 & 0.024 & 0.206 & 0.005 \\
6364\ [O\ {\sc i}]\       & 0.042 & 0.003 & 0.005 & 0.000 \\
6548\ [N\ {\sc ii}]\      & 3.103 & 0.118 & 0.282 & 0.012 \\
6563\ H$\alpha$\          &32.555 & 0.721 & 2.906 & 0.074 \\
6584\ [N\ {\sc ii}]\      & 8.831 & 0.215 & 0.770 & 0.021 \\
6678\ He\ {\sc i}\        & 0.616 & 0.015 & 0.048 & 0.001 \\
6717\ [S\ {\sc ii}]\      & 0.761 & 0.018 & 0.057 & 0.002 \\
6731\ [S\ {\sc ii}]\      & 0.915 & 0.021 & 0.068 & 0.002 \\
7065\ He\ {\sc i}\        & 0.434 & 0.012 & 0.022 & 0.001 \\
7136\ [Ar\ {\sc iii}]\    & 2.117 & 0.051 & 0.102 & 0.003 \\
7237\ [Ar\ {\sc iv}]\     & 0.057 & 0.003 & 0.003 & 0.000 \\
7281\ He\ {\sc i}\        & 0.170 & 0.010 & 0.007 & 0.000 \\
7320\ [O\ {\sc ii}]\      & 0.174 & 0.014 & 0.007 & 0.001 \\
7330\ [O\ {\sc ii}]\      & 0.156 & 0.014 & 0.006 & 0.001 \\[0.2cm]
C(H$\beta$)\ dex          & \MC {4}{c}{3.16$\pm$0.03} \\
$E_{B-V}$\ mag            & \MC {4}{c}{2.15$\pm$0.01} \\
\bottomrule
\end{tabular}
\end{center}
\end{table}

The spectrum of the nebula of Abell\,48
has never been studied before,
because it is relatively faint and highly reddened.
An objective prism spectrum was taken by \citet{San76},
where only the H$\alpha$\ line was detected.
\cite{depew2011} studied the CS of Abell\,48
with the Siding Spring Observatory~2.3m telescope, but they did not publish any spectrum.
Even with SALT we were not able to detect any useful
signal in the spectral range below 4300\,\AA.
Our final 1D spectrum of the nebula is shown in Fig.~\ref{fig:PN_spec}.

Emission lines 
in the spectrum of the surrounding nebula
were measured applying the MIDAS programs described in
detail in \citet{SHOC,Sextans}.
Table~\ref{tab:Intens} lists the measured relative
intensities of all detected emission lines relative to H$\beta$
(F($\lambda$)/F(H$\beta$)), the ratios corrected for the extinction
(I($\lambda$)/I(H$\beta$)), and the derived extinction
coefficient $C$(H$\beta$).
The latter corresponds
to $E_{B-V}$ of 2.15 mag, which is very similar to the value found
independently from the spectrum of the
CS (see Table~\ref{tab:parameters}).

The spectrum of the nebula was interpreted by the technique of 
plasma diagnostics
in the way described in detail in \citet{Ketal08}.
The electron temperatures $T_{\rm e}$(\ion{O}{iii}), $T_{\rm e}$(\ion{N}{ii}),
were calculated directly using the weak auroral lines of oxygen
[O\,{\sc iii}] $\lambda$4363 and nitrogen [N\,{\sc ii}] $\lambda$5755.
Both temperatures and the number density $n_{\rm e}$(\ion{S}{ii}) 
are given in Table~\ref{tab:Chem}.
From the detected emission lines we were able to determine
the total element abundances for O, N, Ar, and He.
These results are shown in Table~\ref{tab:Chem}
together with solar abundances \citep{Asplund09}.

The derived abundances of the $\alpha$-elements O and Ar are lower
than solar by about $-0.3$\,dex. 
In contrast,  He and N 
show abundances close to the solar values.

The IR fluxes at wavelengths longer than 8\,$\mu$m, which have been
measured by {\it Midcourse Space Experiment} ({\it MSX}) satellite
\citep{price2001}
 and the {\it Wide-field Infrared Survey Explorer}
\citep[{\it WISE};][]{wright2010},
show a strong excess compared to the predicted stellar continuum.
This excess could be due to the low resolution of these
instruments, which results in confusion between the stellar
source and the nebular emission.

\section{Discussion}
\label{sect:discussion}

As discussed in the introduction, there were doubts whether Abell\,48
is a PN with a low-mass central star or a ring nebula
around a massive star.
Now we discuss several arguments in favour of the PN status of this object.

\subsection{Abell\,48 -- a PN or a ring nebula around a massive WR star?}
\label{sect:PN_or_WR}

The analysis of the spectrum of the nebula of Abell\,48 gives us 
indications that this object is most likely a PN,
and not a nebula around a massive star:\\
Diagnostic diagrams are very often used to classify emission line
sources and separate them from each other on the base of the most important
line ratios \citep[see e.g.,][and references therein]{KPZ08,FP10}.
The most frequently used diagnostic diagram is
{log({F(H$\alpha$)/F([N\,{\sc ii}] $\lambda\lambda$6548,6584))}
versus
{log({F(H$\alpha$)/F([S\,{\sc ii}] $\lambda\lambda$6717,6731))}.
For Abell\,48 these ratios are 0.44 vs.\ 1.37. 
With these values, Abell\,48 is located in the middle of the region
occupied by PNe and far away from the place where H{\sc\,ii} regions
and WR shells are typically found.

The density of the Abell\,48 nebula and the diagnostic diagram
{log({[S\,{\sc ii}] $\lambda$6717/$\lambda$6731))}
versus
{log({F(H$\alpha$)/F([N\,{\sc ii}] $\lambda\lambda$6548,6584))}
also puts this source into the region of PNe 
(cf.\ Figure~\ref{fig:PN_diag}), 
but not of  WR shells, which all have
densities similar to those of H{\sc\,ii} regions
\citep{Esteban92,Esteban94,GC94,Greiner99,SBW11,FMVPRS12}.

\begin{figure*}
 \includegraphics[width=\textwidth]{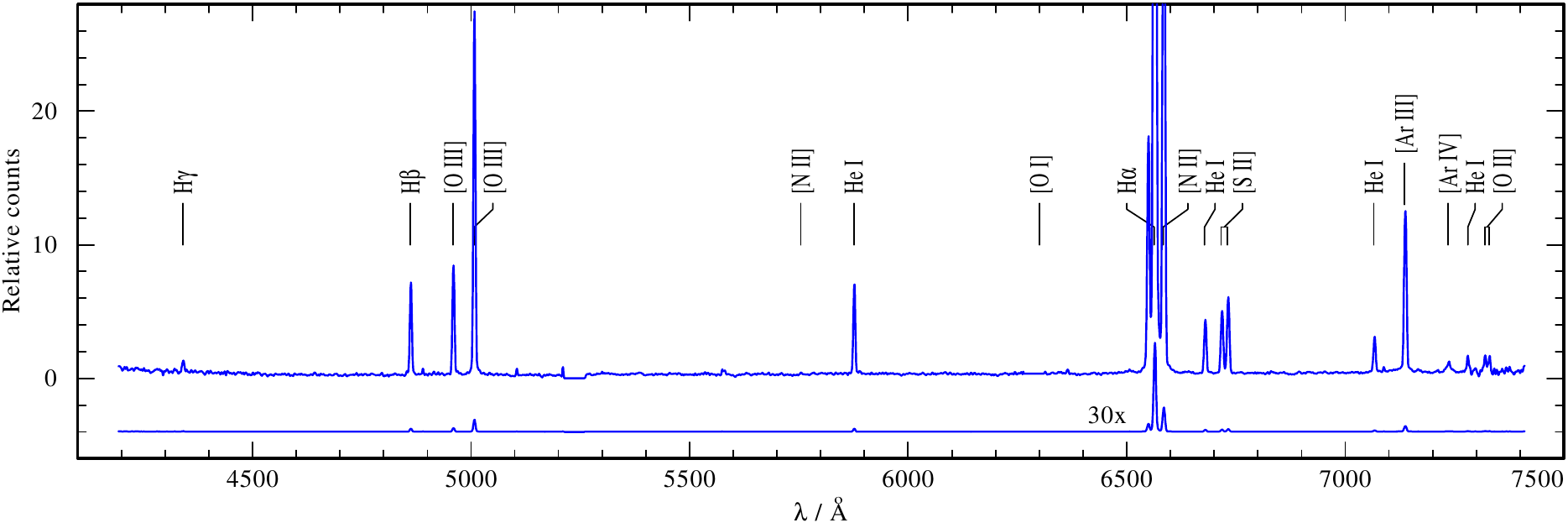}
 \caption{
One-dimensional reduced spectrum of the nebula.
Most of the detected strong emission lines of the nebula are marked.
All detected and measured lines for the nebula are listed in 
Table~\ref{tab:Intens}.
The lower spectrum is scaled by 1/30 to show the relative intensities of 
strong lines. 
\label{fig:PN_spec}}
\end{figure*}

\begin{table}
\begin{center}
\caption{Elemental abundances in the nebula of Abell\,48.}
\label{tab:Chem}
\begin{tabular}{lr @{$\pm$} lc} 
\toprule
Quantity & \multicolumn{2}{c}{Abell\,48}&Sun$^a$\\ 
\midrule
$T_\text{e}$(O{\sc\,III})\,/\,K\ &$11\,870$&$1\,640$&\\
$T_\text{e}$(N{\sc\,II})\,/\,K\ &$7\,200$&$750$&\\
$n_\text{e}$(S{\sc\,II})\,/\,cm$^{-3}$\ &$1\,000$&$130$&\\[0.3cm]
O$^{+}$/H$^{+}$($\times$10$^5$)\     &$19.97$&$5.89$&\\
O$^{++}$/H$^{+}$($\times$10$^5$)\    &$6.56$&$2.74$&\\
O/H($\times$10$^5$)\                 &$26.53$&$6.49$&\\
12+log(O/H)\                         &$8.42$&$0.11$&$8.69$\\[0.3cm]
N$^{+}$/H$^{+}$($\times$10$^7$)\      &$416.9$&$151.6$&\\
ICF(N)\                              &\multicolumn{2}{c}{$1.55$}&\\
N/H($\times$10$^5$)\                 &$6.48$&$2.36$&\\
12+log(N/H)\                         &$7.81$&$0.16$&$7.83$\\
log(N/O)\                            &$-0.61$&$0.19$&$-0.86$\\[0.3cm]
Ar$^{++}$/H$^{+}$($\times$10$^7$)\    &$6.06$&$2.53$&\\
ICF(Ar)\                             &\multicolumn{2}{c}{$1.21$}&\\
Ar/H($\times$10$^7$)\                &$7.33$&$3.05$&\\
12+log(Ar/H)\                        &$5.86$&$0.18$&$6.40$\\
log(Ar/O)\                           &$-2.56$&$0.21$&$-2.29$\\[0.3cm]
He/H\                                &$0.133$&$0.005$&\\
12+log(He/H)\                        &$11.12$&$0.02$&$10.93$\\
\bottomrule
\end{tabular}
\end{center}
$^a$ Solar abundances from \citet{Asplund09}
\end{table}

\subsection{Electron-scattering line wings}

Strong emission lines in WR-type spectra show extended line
wings which can be attributed to line photons that are redistributed by
scattering on free electrons in the wind. The strength of these wings
is sensitive to the degree of wind inhomogeneities, the so-called
clumping (see Section~\ref{sect:fitting}). 
The spectra of massive WN stars are typically
consistent with a clumping density contrast of $D=4\,\ldots\,10$ 
\citep{hillier1991,hamann1998,hamannliermann2006}.

In the spectrum of the CS of Abell\,48 these electron-scattering line
wings are very weak and barely visible, while in contrast  the massive WN
models display significantly stronger wings. 

However, the models which we have calculated for central star
parameters, i.e.\ with $\log L/L_\odot=3.7$ and
$M=0.6\,M_\odot$ (cf.\ Section~\ref{sect:analysis}), show only weak 
electron-scattering line wings with the same
values for the clumping
contrast as for massive WN stars, consistent with the
observation.
We consider this as an evidence for the low-mass nature of the CS of
Abell\,48.

\subsection{Location and luminosity of the CS of Abell\,48}
\label{sect:location}

The best fit to the SED of the CS of Abell\,48 was achieved with a
color excess $E_{B-V} =2.10\pm0.05$ mag (which agrees
well with $E_{B-V} =2.15\pm0.01$ mag derived from the line ratios
in the spectrum of the nebula; see Section~\ref{sect:Phys}),
implying an extinction in the visual of $A_V \approx 7$\,mag. The
latter translates into $A_{K_\text{s}} \approx 0.8$\,mag, if one
adopts the extinction law from \cite{rieke1985}.
These values should be compared with the full
Galactic $V$ and $K_\text{s}$-band extinctions in the direction of
Abell\,48, which according to \cite{schlegel1998}\footnote{See also
\url{http://irsa.ipac.caltech.edu/applications/DUST/}} 
are $\approx 34$
and 4\,mag, respectively. Thus, one can expect that Abell\,48 is
located at a factor of $\sim 5$ smaller distance, $d$, than the
line-of-sight depth of the Galaxy of $\approx 20$\,kpc 
\citep[e.g.][]{churchwell2009},
i.e. at $d\approx 4$\,kpc. Similar
distance estimate ($d\approx 3.9$\,kpc) could be derived if one
assumes an average of 1.8\,mag of visual extinction per kpc in the
Galactic plane. 

These distance estimates are supported by a study of the distribution
of interstellar extinction towards $l\approx 29\degr$, which shows that
$A_{K_\text{s}}$ grows from $\approx$0.3 to 1.0\,mag between 2.5 and 4\,kpc,
then reaches a values of $\approx$1.5 mag at 6\,kpc, and then abruptly
increases to $\approx$2.5\,mag between 6 and 7\,kpc 
\citep{negueruela2011}.
Since $A_{K_\text{s}}$ towards Abell\,48 is $\approx$0.8\,mag, one has that
this object should be located in the Scutum-Centaurus arm at $d<4$\,kpc
\citep[see fig.~12 in][]{negueruela2011}
and that its luminosity of
$\log L/L_{\odot} \leq 4.3$ is much smaller than the lowest plausible
luminosity for a massive WN star of $\log L/L_{\odot} \ga 5.3$
\citep{hamannliermann2006},
but within the range of luminosities
derived for CSPNe 
\citep[e.g.][]{miller-bertolami2007}. 
We adopt the
typical luminosity of CSPNe of $L=6000\,L_{\odot}$ 
(see Section~\ref{sect:analysis}), 
which
leads to a distance to Abell\,48 of $d=1.9$\,kpc. 
Note that the luminosity
of Abell\,48 and its
$R_\ast$ and
$\dot{M}$ 
(given in Table~\ref{tab:parameters}) 
can be scaled
to different distances as $\propto d^2$, 
$\propto d$,
and $\propto d^{3/2}$, respectively 
\citep{schmutz1989}.

Alternatively, if
the CS of Abell 48 were a massive WN star, then it should be located well
beyond the Sagittarius Arm. Indeed, if using the mean $K_\text{s}$-band absolute
magnitude of massive WN5-6 stars from 
\citet{crowther2006},
$M_{K_\text{s}}=-4.41$\,mag and the $K_\text{s}$-band extinction derived from
the SED fitting and line ratios in the spectrum of the nebula, one gets a
distance modulus of 15.94\,mag, which translates into a distance of 15.4\,kpc.
An even larger distance would be derived, if one estimates $A_{K_\text{s}}$
from the intrinsic $J-K_\text{s}$ color of WN5-6 stars of 0.18\,mag
given in
\citet{crowther2006}.
In this case, one gets $A_{K_\text{s}}=0.66$\,mag and
$d=16.4$\,kpc. These two distance estimates imply that Abell\,48 would
be located
either in the Perseus or the Outer Arm. This location, however, 
is in contradiction 
with the moderate extinction towards Abell\,48 in view of the
presence of an extinction wall at $\approx 6$\,kpc 
 \citep[e.g.][]{negueruela2011}.

\begin{figure}
\includegraphics[width=\columnwidth]{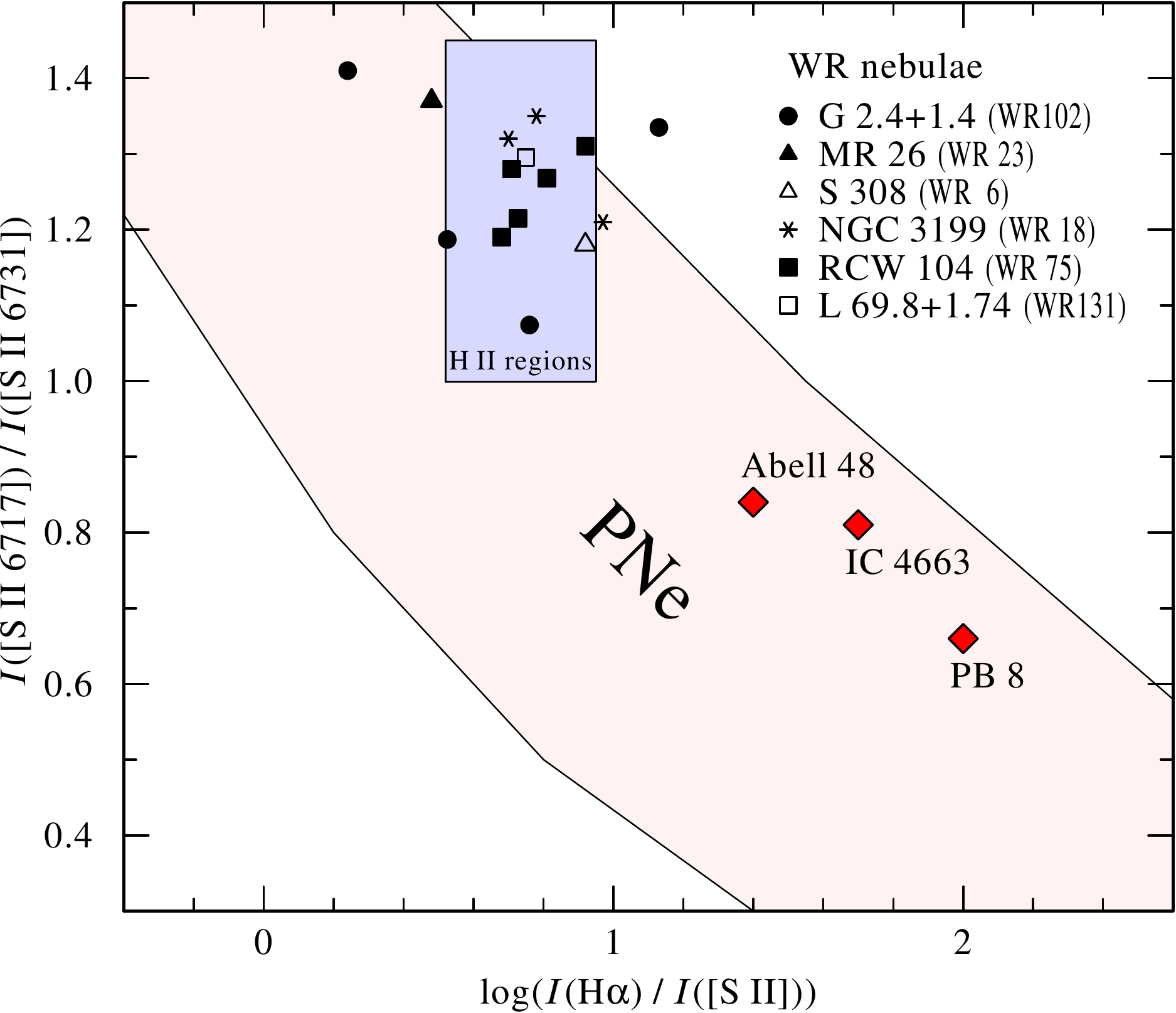}
\caption{ Diagnostic diagram 
H$\alpha$/([S\,{\sc ii}] $\lambda\lambda$6717,\,6731) intensity ratio
versus [S\,{\sc ii}] $\lambda\lambda6717/6731$ ratio as a measure
of electron density versus
excitation for gaseous nebulae. 
Shown are the locations for H\,{\sc ii} regions 
as well as the positions of WR nebulae from \citep{Esteban92}.
Most PNe lie on the indicated strip from \citep{riesgo2002}.
Also labeled are the positions of the [WN] and [WN/WC]
stars Abell\,48, IC\,4663, and 
PB\,8 to demonstrate their CSPNe status.}
\label{fig:PN_diag}
\end{figure}

\subsection{Abell\,48 as a runaway}

\begin{figure}
\begin{center}
\includegraphics[width=\columnwidth]{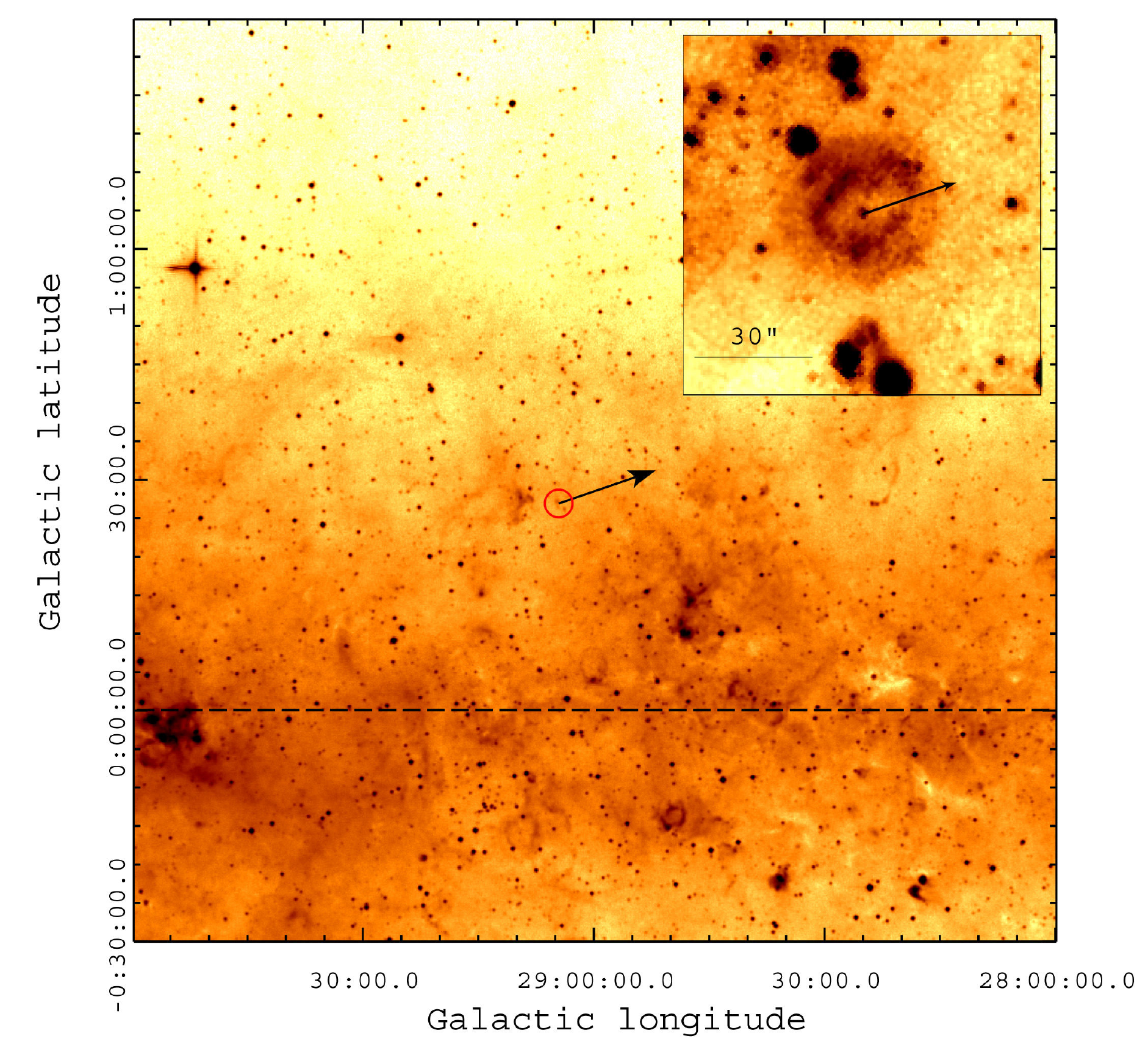}
\end{center}
\caption{$2\degr \times 2\degr$ {\it MSX} 8.3\,$\mu$m image of the
Galactic plane (shown by a dashed line) centered at $l=29\degr,
b=0\fdg5$, with the position of Abell\,48 indicated by a circle.
The inset shows the IRAC 8\,$\mu$m image of Abell\,48. The arrows
on both images show the direction of motion of Abell\,48. The
orientation of the images is the same.
    }
\label{fig:vel}
\end{figure}
The short distance to Abell\,48 is also supported by the detection
of a significant proper motion for its CS: $\mu_\alpha \cos\delta
= -13.3\pm4.8\,\text{mas}\,\text{yr}^{-1}$ and 
$\mu_{\delta} = -14.8\pm4.8\,\text{mas}\,\text{yr}^{-1}$ 
\citep{roeser2010}.
To convert this proper motion and the radial velocity of
Abell\,48 (see Table~\ref{tab:parameters}) into the peculiar
transverse
and radial
velocities, we used the Galactic constants $R_0 = 8.0$\,kpc and
$\Theta _0 =240 \, \text{km} \, \text{s}^{-1}$ 
\citep{reid2009}
and the solar peculiar motion $(U_\odot , V_\odot ,
W_\odot)=(11.1, 12.2, 7.3) \, \text{km} \, \text{s}^{-1}$
\citep{schoenrich2010}.
The derived velocity
components and the total space velocity
are given in Table~\ref{tab:vpec} and 
imply that Abell\,48 is a runaway system
\citep[e.g.][]{blaauw1961}.
For the error calculation, only the errors of the proper motion
and radial velocity
measurements were considered. The peculiar radial velocity of
Abell\,48 contributes only slightly to its total space velocity,
$v_\text{pec}$ (Table~\ref{tab:vpec}).
These estimates show that
the larger $d$ the larger $v_\text{pec}$ and correspondingly the
lower the likelihood that the CS of Abell\,48 can be accelerated
to such high velocity. Consequently, one can conclude that the
short distance to Abell\,48 is more likely.

The runaway status of Abell\,48 is consistent with the orientation
of its peculiar velocity with respect to the Galactic plane.
Fig.\,\ref{fig:vel} shows the
{\it MSX} 8.3\,$\mu$m image of a
$2\degr \times 2\degr$ region of the Galactic plane (centered at
$l=29\degr, b=0\fdg5$). One can see that Abell\,48 is moving in
the ``correct" direction, i.e. away from the Galactic plane. On
the other hand, the large margin of error in $v_b$ leaves the
possibility that Abell\,48 is moving almost parallel to the
Galactic plane, which in turn is consistent with the small
height above the Galactic plane (cf.\ Table~\ref{tab:vpec}).

\addtolength{\tabcolsep}{-1pt}
\begin{table}
\begin{center}
\caption{Abell\,48: 
Peculiar transverse (in Galactic coordinates), radial and total space
velocities, and height over the Galactic plane 
for different adopted distances.
}
\label{tab:vpec}
\begin{tabular}{D{.}{.}{1}r@{$\pm$}lr@{$\pm$}lr@{$\pm$}lr@{$\pm$}lr}
\toprule
\multicolumn{1}{c}{$d$}   
& \multicolumn{2}{c}{$v_l$}          
& \multicolumn{2}{c}{$v_b$} 
& \multicolumn{2}{c}{$v_\text{r}$}    
& \multicolumn{2}{c}{$v_\text{pec}$} 
& \multicolumn{1}{c}{$d\sin b$}
\\
\multicolumn{1}{c}{[kpc]} 
& \multicolumn{2}{c}{[km\,s$^{-1}$]} 
& \multicolumn{2}{c}{[km\,s$^{-1}$]} 
& \multicolumn{2}{c}{[km\,s$^{-1}$]} 
& \multicolumn{2}{c}{[km\,s$^{-1}$]}   
& \multicolumn{1}{c}{[pc]} 
\\
\midrule
 1.9 & $-149$&$43$  & $53$&$43$   & 37&4 & $162$&$42$   & 15\\
 4.0 & $-296$&$91$  & $104$&$91$  & -8&4 & $314$&$91$  & 32\\
15.4 & $-970$&$350$ & $378$&$350$ & 74&4 & $1043$&$349$  & 122\\
\bottomrule
\end{tabular}
\end{center}
\end{table}
\addtolength{\tabcolsep}{+1pt}

\subsection{Classification}

Based on the arguments above we conclude that Abell\,48 is a PN
with a low-mass central star.

The abundance pattern, that our spectral analysis revealed, 
corresponds to a WN-type composition,
where is helium dominates, nitrogen is enhanced and the carbon and
oxygen abundances are roughly solar.
Therefore, we suggest to assign Abell\,48 to the [WN] class,
becoming thus the second known member of this spectral class
after IC\,4663 \citep{miszalski2012}.

To determine the detailed subtype of Abell\,48, 
we apply the 3D classification
scheme by \citet{smith1996}. The measured line ratios
(see Table~\ref{tab:abell48-peakratios} and \ref{tab:subtype}) 
suggest that the star belongs to the [WN\,5] subtype.

In this connection, we note that none of the massive WNE stars
possess a compact circumstellar nebula 
\citep[][and references therein]{gvaramadze2009,gvaramadzefabrika2010a},
which provides one
more evidence in support of the PN status of Abell\,48.

\begin{table}
\begin{center}
\caption{Measured line ratios from the spectrum of
the CS of Abell\,48 for spectral subtype classification.}
\label{tab:subtype}
\begin{tabular}{lc}
\toprule
line ratio & peak/continuum \\
\midrule
 N{\,\sc  v}\,4604\,/\,N{\,\sc iii}\,4640 & 2.0\\
 C{\,\sc iv}\,5808\,/\,He{\,\sc ii}\,5411 & 0.7\\
 C{\,\sc iv}\,5808\,/\,He{\,\sc  i}\,5878 & 1.7\\
\bottomrule
\end{tabular}
\end{center}
\end{table}

\subsection{Evolutionary status}

Spectroscopically, Abell\,48 with its strong helium and nitrogen
emission lines belongs clearly to the 
[WN] class. However, Abell\,48 seems to be different in some
respect from the other [WN] star IC\,4663.
While \cite{miszalski2012} could not find
any residual hydrogen in IC\,4663, 
the presence of some hydrogen is indicated by our optical spectrum
of the CS of Abell\,48.
Such residual hydrogen would rule out
a merger scenario with a sequence He-WD + He-WD $\rightarrow$ [WN] 
$\rightarrow$ O(He) as proposed by Reindl et al.\ (in prep.) 
for the O(He) and [WN] stars.
Moreover, if the nebula of Abell\,48 has been ejected during the
merging process it would be hydrogen-deficient, contrary to the
observation. 
Thus, there must exist different evolutionary channels for the
formation of these two [WN] stars.

For the [WN/WC] star PB\,8 \cite{millerbertolami2011} suggested a 
diffusion-induced nova (DIN) of a low-mass ($M \lesssim 0.6 \, M_\odot$)
post-AGB object as the origin of such stars. Among the models calculated for the
different scenarios of the DIN there is one which includes overshooting 
in the convective zone generated during the CNO-flash, which has
abundances that are very similar to those that we found for Abell\,48.
The surface abundances of the model are H:He:C:N:O=22:73:0.66:3.8:0.1
per cent by mass, that is close to our results of
H:He:C:N:O=10:85:0.3:5:0.6 for the wind of Abell\,48. 

The nebula abundance ratio N\,/\,O = 0.25 is below the sharp limit of
N\,/\,O$>0.8$ that \cite{kaler1989} deduced for N-enriched PNe, which
are supposed to indicate hot bottom burning in more massive AGB-stars with
$M_\text{core}> 0.8\,M_\odot$. The low core mass of Abell\,48
would be in line with the models by \cite{millerbertolami2011}.

However, the stellar evolutionary models by \cite{millerbertolami2011}
assume a low-metallicity ($Z=0.001$) progenitor, which is at least for
the CS of PB\,8 not the case, as we inferred solar-like iron-group abundances
from  the UV spectra of this object. For the CS of Abell\,48 we could not
determine any iron-group abundances, but the metallicity of the
nebula is roughly solar (cf.\ Table~\ref{tab:Chem}).

Further problems for this scenario are posed by the involved
timescales. Assuming a PN radius of 0.2\,pc as deduced in 
Section~\ref{txt:PV_body} and  
a typical expansion velocity of $30\kms$, the nebula
age is only 6500\,yr. This is far too short compared to the
evolutionary time
since the ejection of the nebula at the AGB
\citep[$10^6-10^7$\,yr,][]{millerbertolami2011},  
especially since this scenario works only for low stellar masses $M
\lesssim 0.6 \, M_\odot$.

\section*{Acknowledgments}

All SAAO and SALT co-authors acknowledge the support
from the National Research Foundation (NRF) of South Africa.
We would also like to thank Martin Wendt for fruitful discussions of
$\chi^2_\nu$-fitting techniques.

This work has made use of the NASA/IPAC Infrared 
Science Archive, which is operated by the Jet 
Propulsion Laboratory, California Institute of 
Technology, under contract with the National 
Aeronautics and Space Administration, the SIMBAD
database and the VizieR catalogue access tool, 
both operated at CDS, Strasbourg, France.

\bibliographystyle{mn2e}
\bibliography{abell48}

\label{lastpage}

\end{document}